\documentclass[journal, onecolumn]{IEEEtran} 
\usepackage{cite} 
\usepackage{amsmath} 
\usepackage{upgreek,color,url}
\usepackage{graphicx} 
\usepackage{physics} 
\usepackage{caption}
\usepackage{subcaption} 
\usepackage{float}
\usepackage{soul}

\begin{document} \title{Digital Doppler-cancellation servo for ultra-stable
optical frequency dissemination over fiber}
\author{Shambo~Mukherjee, Jacques~Millo,
Baptiste~Marechal, S\'everine~Denis, Gwenha\"el~Goavec-M\'erou, Jean-Michel~Friedt, Yann~Kersal\'e,
Cl\'ement~Lacro\^ute\thanks{The authors are with the FEMTO-ST Institute,
univ. Bourgogne Franche-Comt\'e, CNRS, ENSMM, 26 rue de l’\'Epitaphe, 25000
Besan\c{c}on, France (e-mail: clement.lacroute@femto-st.fr).}}


\pagenumbering{roman} \maketitle

\begin{abstract} Progress made in optical references, including ultra-stable Fabry-Perot cavities, optical frequency combs and optical atomic clocks, have driven the need for ultra-stable optical fiber networks. Telecom-wavelength ultra-pure optical signal transport has been demonstrated on distances ranging from the laboratory scale to the continental scale. In this manuscript, we present a Doppler-cancellation setup based on a digital phase-locked loop for ultra-stable optical signal dissemination over fiber. The optical phase stabilization setup is based on a usual heterodyne Michelson-interferometer setup, while \textcolor{black}{the Software Defined Radio (SDR) implementation of the phase-locked loop is based on} a compact commercial board embedding a field programmable gate array, analog-to-digital and digital-to-analog converters. Using three different configurations including an undersampling method, we demonstrate a \textcolor{black}{20 m long} fiber link with residual fractional frequency instability as low as $10^{-18}$ at 1000~s, and an optical phase noise of $-70$~dBc/Hz at 1 Hz with a telecom frequency carrier.
\end{abstract}
 
\section{Introduction} 

With the joint progress of ultra-stable Fabry-Perot (FP)
cavities, optical frequency combs (OFCs) and atomic optical standards, the
development of optical frequency references has reached both an unprecedented
level of performances and an advanced degree of industrial integration. While
laboratory FP cavities have reached the $10^{-17}$ level \textcolor{black}{\cite{Haefner2015,
Matei2017, Zhang2017, Milner2019, Robinson2019}}, many transportable prototypes have been developed
\cite{Leibrandt2011, Vogt2011, Argence2012, Haefner2020} and several companies
have developed FP-based ultra-stable lasers. OFCs have also become a wide-spread
laboratory tool, and transportable models have been demonstrated
\cite{Lezius2016}. Optical atomic clocks have shown the same level of progress,
with both optical lattice and single-ion clocks reaching the $10^{-18}$ accuracy
level and below \cite{Hinkley2013, Bloom2014, Ushijima2015, Huntemann2016,
Brewer2019}.

This progress in ultra-stable optical references impacts many fields, including
time and frequency metrology, precision spectroscopy, microwave photonics
\cite{Giunta2020} and fundamental physics \cite{Derevianko2014, Safronova2018}.
Several proposals have argued for the development of ultra-stable optical fiber
networks on the continental scale \cite{Riehle2017, Lisdat2016}. Such networks
are already being deployed within national and international frameworks. In
practice, ultra-stable optical links have been implemented at several scales,
from the laboratory or campus \textcolor{black}{\cite{Kudeyarov2018, Beloy2021}} to inner- and inter-city
links \textcolor{black}{\cite{Newbury2007, Bercy2014b, Cantin2021}} and beyond, with long-haul fiber links
\textcolor{black}{\cite{Lopez2012, Calonico2014a, Lopez2015, Clivati2020}}.

\begin{figure*}[h!] \centering
\includegraphics[width=\textwidth]{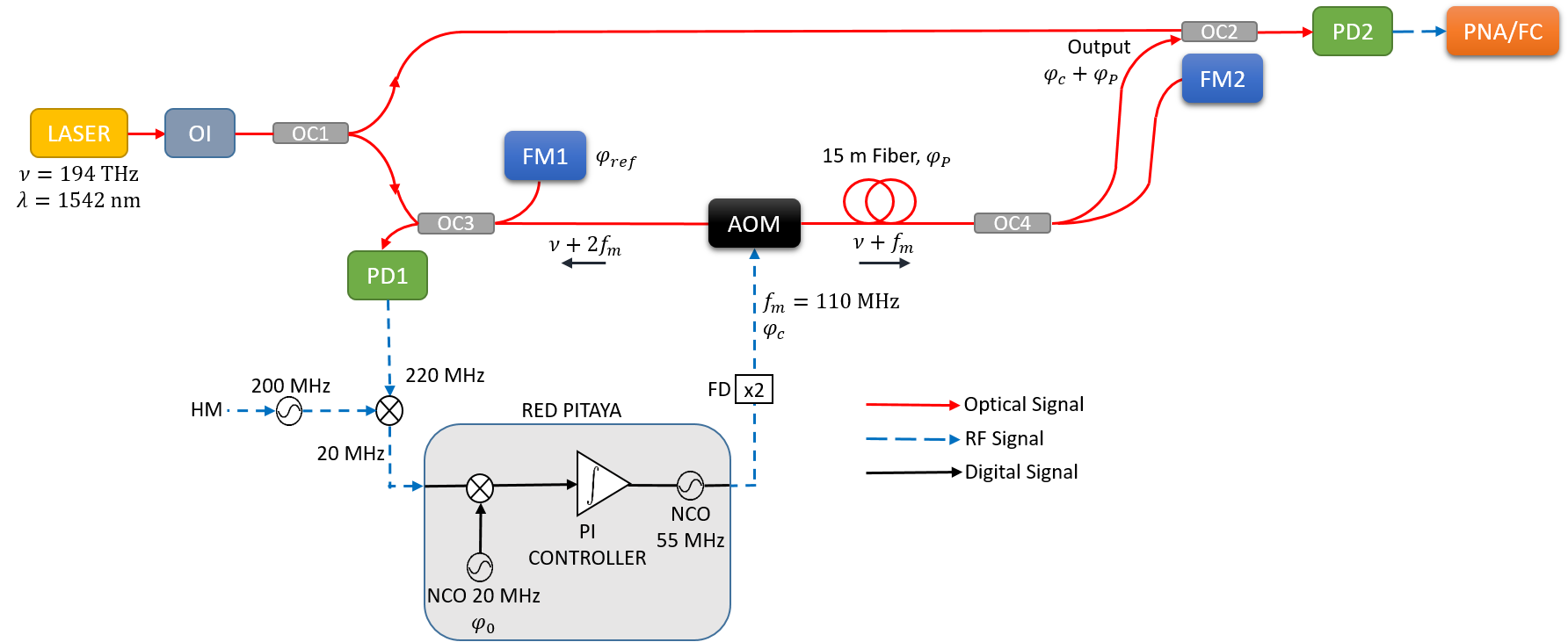}
    \caption{Experimental scheme of the stabilized optical signal transfer through a 20~m long optical fiber link. The stabilization setup consists in a heterodyne Michelson interferometer (HMI) formed between optical coupler OC3 and the Faraday mirrors FM1 and FM2. Digital control of the optical phase measured at
photodiode 1 (PD1) is performed through a digital phase-locked loop. Characterization of the link output phase noise is performed by comparison to the laser source through a Mach-Zehnder interferometer formed between optical couplers OC1 and OC2. Relevant frequencies and phases have been indicated. See text for details. HM: Hydrogen Maser, OI: Optical Isolator, OC: Optical Coupler, PD: Photodiode, AOM: Acousto Optic Modulator, PLL: Phase-Locked Loop, FM: Faraday Mirror, NCO: Numerically Controlled Oscillator.}
\label{fig:setup} \end{figure*}

Such ultra-stable fiber links are based on a \textcolor{black}{heterodyne} Michelson interferometer in which
one of the arms is shortened to exhibit length fluctuations as small as
possible. The second arm includes the required length to transfer the optical
signal to the end-user. The resulting \textcolor{black}{beatnote phase fluctuations are} directly
related to the length difference \textcolor{black}{fluctuations} between the reference arm and the optical fiber
link. If the length of the reference arm is stable enough, phase fluctuations
detected with a photodiode at the interferometer output corresponds to the link
optical length fluctuations caused by vibrations and thermal perturbations. From
this signal an active loop filter generates a compensation signal to cancel the link phase
noise. This is the so-called Doppler-cancellation scheme \cite{Ma1994, Ye2003,
Newbury2007}, which can also work by post-processing the link phase fluctuations
\cite{Calosso2015}. State-of-the-art \textcolor{black}{long-haul optical fiber links} reach a residual link
fractional frequency stability in the $10^{-19}$ or below \textcolor{black}{for integration times above 1000~s seconds}, on
distances of over 1000 km \cite{Calonico2014a, Stefani2015, Raupach2015,
Chiodo2015a}.

In parallel to such advances, the recent years have seen the advent of
digital-electronics in the fields of stabilized lasers, time-and-frequency
metrology and atomic physics. With analog-to-digital and digital-to-analog
converters (ADCs and DACs),  field programmable gate arrays (FPGAs) and direct
digital synthesis (DDSs) having reached levels of performance, integration and
ease-of-use competitive with analog electronics including for radiofrequency (RF) applications, several laboratories and
companies have started to move from analog to digital electronics control loops
\cite{Bowler2013, Huang2014, Pruttivarasin2015, Stuart2019}. Such Software Defined Radio (SDR) setups bring the possibility to implement different schemes using the same hardware with the benefit of software long term stability and reconfigurability when implementing classical analog processing techniques, and enable the simple duplication of systems and electronic functions, the possibility to control many different channels in parallel with a high data rate and strong computing power, along with small components footprint. Beyond these advantages, discrete time digital signal processing brings a new paradigm illustrated here with sub-sampling and benefiting from high order Nyquist zones \cite{sdra}. Digital implementation of the local oscillator as a numerically controlled oscillator (NCO), and mixing in the FPGA removes the classical analog challenges of local oscillator leakage through the mixer and I,Q imbalance when recovering the phase from a quadrature demodulation. This approach has led to numerous novel developments
\cite{hasselwander2016gr,sherman2016oscillator,mahnke2018characterization,balakirev2019resonant,tourigny2018open,stimpson2019open,preuschoff2020digital}.

Among these approaches, Tourigny-Plante \textit{et al.} use the CORDIC arctan algorithm to digitally extract the signal phase \cite{tourigny2018open}. The phase-locked loop (PLL) is implemented on the same board used in this manuscript, clocked by its internal quartz, using an open-source code available online \cite{Deschenes}. The digital PLL is demonstrated on a Doppler-cancellation scheme similar to the one presented here and characterized in-loop. Another approach is presented by C\'ardenas-Olaya \textit{et al.}, using a tracking DDS scheme that is, to our knowledge, not openly documented \cite{Olaya2021}. The noise of the digital instrumentation has been carefully characterized \cite{Olaya2016} and demonstrated experimentally by two-way compensation of long-haul optical fiber links \cite{Olaya2019a}.

In this manuscript, we present a digitally controlled Doppler-cancellation setup developed to provide a stabilized fiber output to a compact FP cavity \cite{Didier2018} on the
laboratory scale. We have put the emphasis on simplicity using off-the-shelf
optical components and a compact, digital electronics control loop. \textcolor{black}{The digital control loop can be fully reproduced using the OSC-IMP environment, with all code being openly released under the CeCILL Licence \cite{cecil}. As presented in the next Section, we rely on a simple demodulation scheme to access the signal phase noise, different from the arctan approach developed in \cite{tourigny2018open} or the tracking DDS approach of \cite{Olaya2021}. While in-situ characterization of the Ref. \cite{Olaya2016} setup is implemented using two-way compensation \cite{Olaya2019a}, we present a more traditional Doppler-compensation setup with out-of-loop characterization. In addition to the standalone configuration of the digital board, presented also in \cite{tourigny2018open}, we characterize external referencing and undersampling using this board.}

Section \ref{sec:methods} presents the setup, including both
the optical interferometers and the control electronics. Section
\ref{sec:results} illustrates the results obtained in three different
configurations, namely using a stand-alone digital board, an externally clocked
digital board, and using undersampling. Section \ref{sec:discussion} puts these
results in perspective with the current state-of-the-art and applications of
ultra-pure signal optical fiber transfer.

\section{Methods} \label{sec:methods} 

The aim of the setup is to transfer an
ultra-stable signal at the laboratory scale (few tens of meters) using an
optical fiber while maintaining the spectral purity and frequency stability of
the signal. To do so, the main challenge is to reduce the phase disturbances
caused by the thermal sensitivity and mechanical vibrations sensitivity of the
optical fiber through the path of the link and to compensate these phase
fluctuations using a PLL.

\subsection{Principle of the compensated link} \label{sec:ppe}

The link, including the noise compensation, is based on a heterodyne Michelson
interferometer (HMI) scheme (Fig. \ref{fig:setup}) in which the beam splitter is
the fiber coupler OC3. The short arm of the Michelson interferometer that
defines the phase reference $\varphi_{ref}$ of the link is formed between OC3
and the Faraday mirror FM1. The long arm used to transfer the optical reference
that includes the 15~m of optical fiber is the path between OC3 and another
Faraday mirror FM2 in which the phase is affected by environmental perturbations
$\varphi_{P}$.  This path includes an acousto-optical modulator (AOM) driven at
$f_{m}\approx~110~\mathrm{MHz}$ that adds a controlled phase $\varphi_{c}$ to
this arm. The optical coupler OC4 extracts the signal at the end of the link,
with phase $\varphi_{u}=\varphi_{P}+\varphi_{c}$. The beatnote signal on PD1
carries the phase of the link at $2f_{m} \simeq 220 \ \mathrm{MHz}$ and is
demodulated by an RF reference signal at $f_{0}=220\
\mathrm{MHz}$\footnote{Please note that in practice we use two different
schemes, with cascaded analog and digital demodulation at 200~MHz and 20~MHz or
direct digital demodulation of the 220~MHz beatnote.}. The error signal
$\varepsilon =2\left(\varphi_{P} + \varphi_{c}-\varphi_{ref}
\right)-\varphi_{0}$ with $\varphi_{0}$ the contribution of this RF reference,
vanishes when the loop filter is activated to compensate the noise of the link.
Please note that the factor $2$ comes from the double pass of the light in the
two arms of the HMI. Thus the phase at the output of the link is $\varphi_{u} =
\varphi_{ref} + \frac{1}{2} \varphi_{0}$. With this basic model we see that the
noise compensation is ultimately limited by the phase fluctuations of the
reference arm $\varphi_{ref}$ and the RF reference $\varphi_{0}$. \textcolor{black}{A more detailed analysis of the link various noise contributions can be found in \cite{williams2008}}.

\subsection{Optical setup}

The experiment is fed by a \textcolor{black}{free-running}, commercial fibre laser (NKT Photonics Koheras
\textcolor{black}{Adjustik, with a typical phase noise of $80\ \rm{dBrad}^2/$Hz at 1~Hz}) of wavelength $\lambda=1542 \ \mathrm{nm}$ with about $14\
\mathrm{mW}$ \textcolor{black}{optical output power}. The 90/10 fiber beam splitter OC1 splits the laser into the
two parts of the setup, with 10\% going to the characterization part. The light
is injected in the \textcolor{black}{15~m fiber} through the 90/10 coupler OC3 using the 90\%-port.
At the end of the link the 90/10 coupler OC4 sends 90\% to the output port
and 10\% to Faraday Mirror FM2. With this configuration 35\% of the power
injected in the link is transmitted to the end user (at the output point on Fig.
\ref{fig:setup}) and the beatnote signal on PD1 is realized with 8.1\% of the
laser output power coming from the reference arm and 0.17\% from the round trip.
\textcolor{black}{Should this be too low,} this could be increased to 1.3\% by using 50/50 couplers
for OC3 and OC4, with an end user total transmission decreased to 11\%. 

Since the signal provided by the photodiode PD1 is processed by the PLL to phase
lock the link (in-loop signal), it can not be used to evaluate the performances
of the link. A dedicated part of the setup is implemented to characterize the
signal transmitted by the link (Fig. \ref{fig:setup}). We use a heterodyne
Mach-Zehnder interferometer (HMZI) formed between beamsplitters OC1 and OC2 to
characterize the link output phase noise. One arm of the HMZI is formed by a
short uncompensated fiber path between OC1 and OC2, while the second arm is
formed by the ultra-stable link and its output fiber.

\textcolor{black}{In practice, the evaluation of performances of the link requires to place the
end close to the beginning. In order to passively reduce the effect of
uncompensated fiber paths (OC1-OC2, OC1-OC3, OC3-FM1 and OC4-OC2, with a total length of 4.5~m), the entire
setup except for the 15~m patch-cord is placed on a floated optical table and
underneath a wooden box that provide passive acoustic isolation and thermal
shielding from the room air conditioning residual temperature fluctuations.}

\textcolor{black}{Moreover, the uncompensated fiber path between OC4 and
OC2 is partially corrected for by placing the FM2 close to OC2
and using the same length of fiber, so that thermal perturbation, fluctuating slowly,
affect these two fibers similarly and the compensation matches the perturbations
in the output fiber.}

The ultra-stable link performance is analyzed by characterizing the $110\
\mathrm{MHz}$ beatnote signal on PD2 using either a phase noise
analyzer (PNA, Rohde $\&$ Schwarz FSWP) or a frequency
counter (FC, K+K FXE) to quantify both phase noise and fractional
frequency stability.

\subsection{Control electronics} 

\begin{figure}[h!] \centering
\includegraphics[width=\columnwidth]{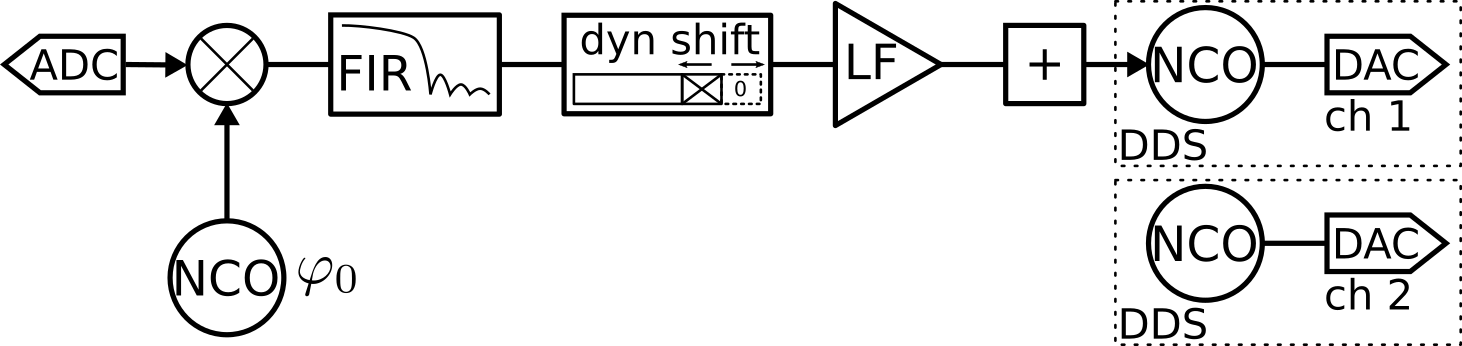}
\caption{Digital control scheme. The digital board offers two parallel channels. The two channels are similar and benefit from the digital implementation of duplicating functionality, emphasizing the flexibility of the \textcolor{black}{SDR} approach. We use the first one for locking. The second channel is used as a DDS for the experiments presented here. The design is based on separate blocks each representing a function. ADC: analog to digital converter. NCO: numerically controlled oscillator. FIR: finite impulse response filter. dyn shift: dynamic shifter (adds or removes least significant bit(s), set on-the-fly by user). LF: loop filter (proportional integrator function). See text for details.} \label{fig:electronics}
\small Details and designs can be found here: \url{https://github.com/oscimp/app/tree/master/redpitaya/double_iq_pid_vco}
\end{figure}

Our control electronics is centered on a digital PLL implemented on an FPGA integrated on a system on a chip (SoC). The commercial board (RP, Red Pitaya STEMlab 125-14) includes an analog to digital converter (ADC, 14 bits, 125 MS/s/channel, two channels), a digital to analog converter (DAC, 14 bits, 125 MS/s/channel, two channels) as well as the SoC including an embedded processor (Xilinx Zynq 7010). \textcolor{black}{Fig. \ref{fig:electronics}} sketches the electronics setup used in the two first measurements of
this work. Modifications for the last step will be described in section \ref{sec:unders}. Compliance with the Nyquist–Shannon theorem during sampling by the ADC, in addition to the front-end filters, requires downconverting the $220\ \mathrm{MHz}$ of the HMI beatnote signal to $20\ \mathrm{MHz}$, below the Nyquist  frequency $f_N=62.5\ \mathrm{MHz}$ of the ADC thanks to a $200\ \mathrm{MHz}$ signal source that is derived from an active hydrogen MASER (HM) using a DDS synthesizer. The resulting signal is filtered and amplified to provide a $11.6\ \mathrm{dBm}$ signal to the input connector of the board taking care to add a $50\ \Omega$ load in parallel to the high-impedance ADC input (the jumper on the board is positioned to the $\pm 1\ \mathrm{V}$ full scale).

The functions to process the signal for digital PLL are the same than with analog electronic: demodulation with a phase reference, low-pass filtering, loop filter and modulation of a controlled oscillator (see Fig. \ref{fig:electronics}). Data provided by the 14-bits ADC, processed by these functions and sent to the DAC are signed and fed with a constant data rate of 125~MS/s. A 16-bits amplitude and 40-bits phase accumulation register NCO is the phase reference of the phase-locked loop. The demodulation is implemented with a multiplication process with a 14 bits output followed by a finite impulse response (FIR) filter configured with 25 coefficients each encoded in 16 bits.

In this manner, the signal recorded from the ADC at baseband is frequency transposed by the multiplication with the complex NCO acting as an ideal mixer that avoids the classical imbalance and lack of quadrature issues found in hardware implementations. The phase detector is a complex to real part converter since no gain was observed when using the more time and resource consuming arctan CORDIC implementation which would extend the linearity range to large phase offsets. In this short link investigation, no phase slip was observed once the lock is activated, and the setpoint remains in the locally linear behaviour of the complex to real phase detector. Using the CORDIC implementation might be necessary for implementing longer links, to ensure linearity and avoid cycle slips.

The FIR filter cutoff frequency is $4\ \mathrm{MHz}$ with more than $40\ \mathrm{dB}$ rejection for frequencies above $14\ \mathrm{MHz}$. The FIR output is a 32 bits data stream. A dynamic  shifter  is  then  used to reduce the number of bits to 14 before the loop filter while adapting the range to the signal dynamics on the fly for the generation of the correction signal. The output is added to a constant value used to bias the frequency of the output DDS based on a second NCO. Then, data are sent to the channel 1 of the DAC. As the Nyquist-Shannon sampling theorem must also be satisfied for this converter, the NCO center frequency is set at $55\ \mathrm{MHz}$ and then the signal is frequency doubled, filtered and amplified to drive the AOM, which then corrects the optical fiber link phase fluctuations. Another DDS is implemented in the FPGA-chip based on a third NCO that outputs to channel 2 of the DAC. For most measurements presented below, this allows us to determine the digital phase reference contribution to the lock performance. The NCOs are implemented as 40 bits accumulators clocked at a rate of
125~MHz, leading to a $125\cdot 10^6/2^{40}\simeq 0.11$ mHz resolution. The accumulator feeds a lookup table generating a sine wave encoded with 4096 values matching the 12-bit DAC resolution.

The embedded processor is used to run a GNU/Linux operating system (OS) dedicated to these applications (Buildroot). A digital signal processing framework dedicated to time \& frequency metrology application has been developed and is available to the community at \url{https://github.com/oscimp/oscimpDigital} providing a consistent set of the real time signal processing blocks configuring the FPGA, the Linux drivers communicating with hardware, and the userspace libraries providing the flexibility of a user friendly interface for controlling the various processing blocks. This OS interfaces most of the basic functions described above, including network connectivity and file storage for long term logging as required for metrology application, and allows modifications of parameters by the users such as filter coefficients and DDS's frequencies. \textcolor{black}{Among other features, a key one is the optimization of the resolution in agreement with the range of the error signal by using the dynamic shift register. This block selects 14 consecutive bits among the data 32 bits. The number of discarded lowest significant bits $n$ is set by the user (where 18-$n$ will be the number of discarded most significant bits). This reduction of the data size can be crucial since each single mathematical operation enlarges the number of bits required to losslessly encode the output.}

\textcolor{black}{Such an inflation of the number of bits in the data flow is not relevant because (i) it will rapidly surpass resources available in the FPGA, (ii) part of the lowest significant bits can often be neglected and (iii) the output DAC ultimately limits the resolution (14 bits for the RP).}

\textcolor{black}{This is also useful for adjusting the resolution when the input level range is changing. A high $n$ will ensure that higher amplitude signals full dynamic is captured without overflows at the servo input, which is useful for monitoring the signal in open-loop (monitoring not shown in the circuit diagram), while a lower $n$ guarantees a full resolution of the nearly vanishing error signal when the PLL is engaged.}

As mentioned above, we use the characterization photodiode (PD2) to analyze the link output phase noise and fractional frequency stability carried by the 110~MHz beatnote. While the PNA input can directly analyze the 110~MHz signal phase-noise, the frequency counter input bandwidth is 60~MHz. When measuring fractional frequency stability, we thus mix the 110~MHz signal from PD2 with a 100~MHz signal from a synthesizer clocked by an \textcolor{black}{HM} in order to generate a 10~MHz signal.

\section{Results} \label{sec:results}

We present the results obtained with our setup, including both phase noise and fractional frequency stability. We have investigated three configurations based on the Figs. \ref{fig:setup} and \ref{fig:electronics} general scheme. In §\ref{sec:RPint}, we use a stand-alone board, and all the digital electronics controls are clocked by the board internal quartz oscillator. In §\ref{sec:RPext}, we use the same scheme but this time adding a highly stable external clock to the digital electronics board. In §\ref{sec:unders}, the Fig. \ref{fig:setup} scheme is slightly modified as we implement undersampling, again using an external clock.

\subsection{Using the board internal quartz oscillator} \label{sec:RPint} 

\begin{figure}[h!] 
\centering \begin{subfigure}{.5\textwidth} 
\centering
\includegraphics[width=8cm]{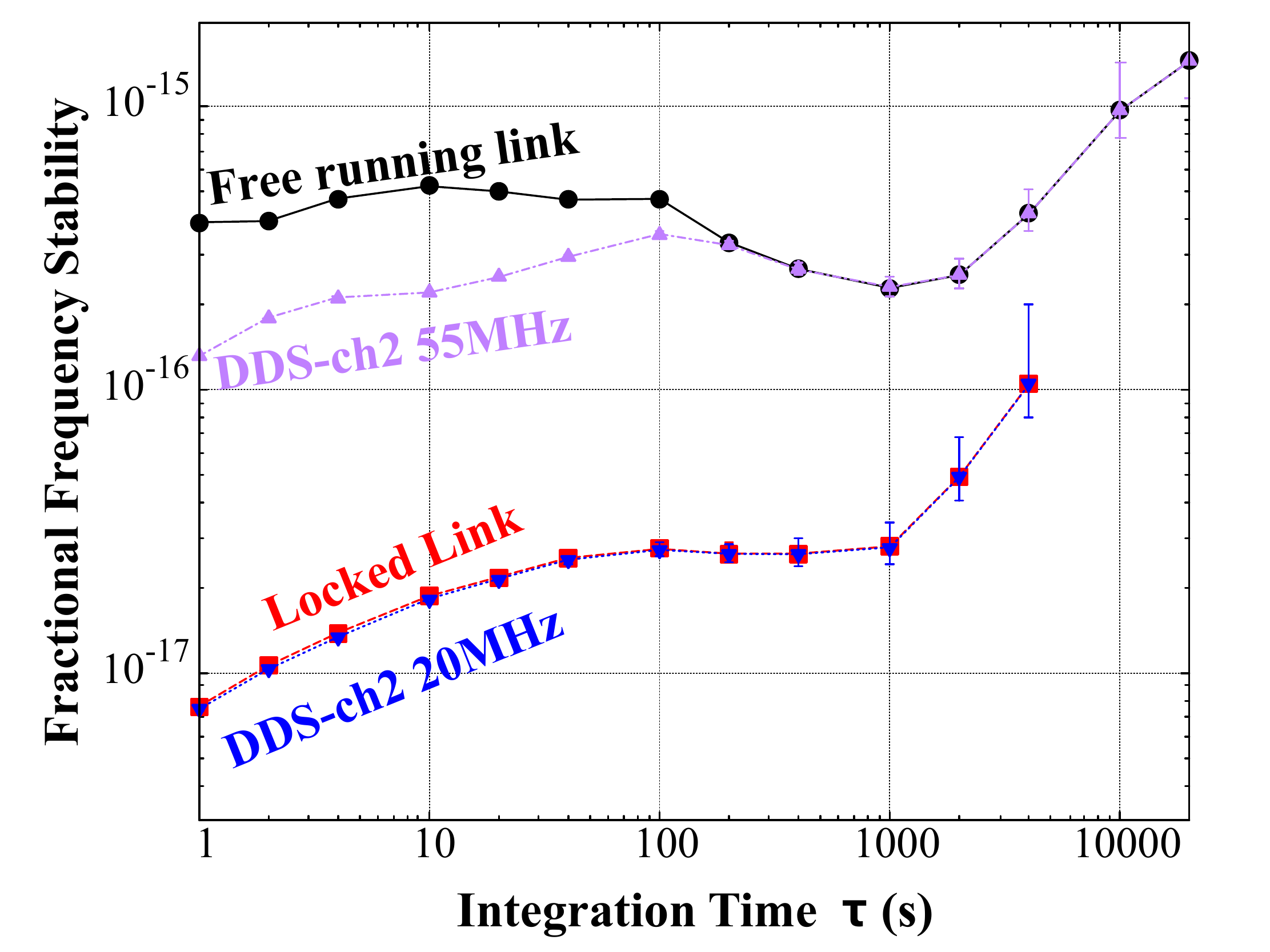} 
\caption{}
\label{fig:FFS of Internal Quartz RedPitaya} 
\end{subfigure}%
\\
\begin{subfigure}{.5\textwidth} \centering
\includegraphics[width=8cm]{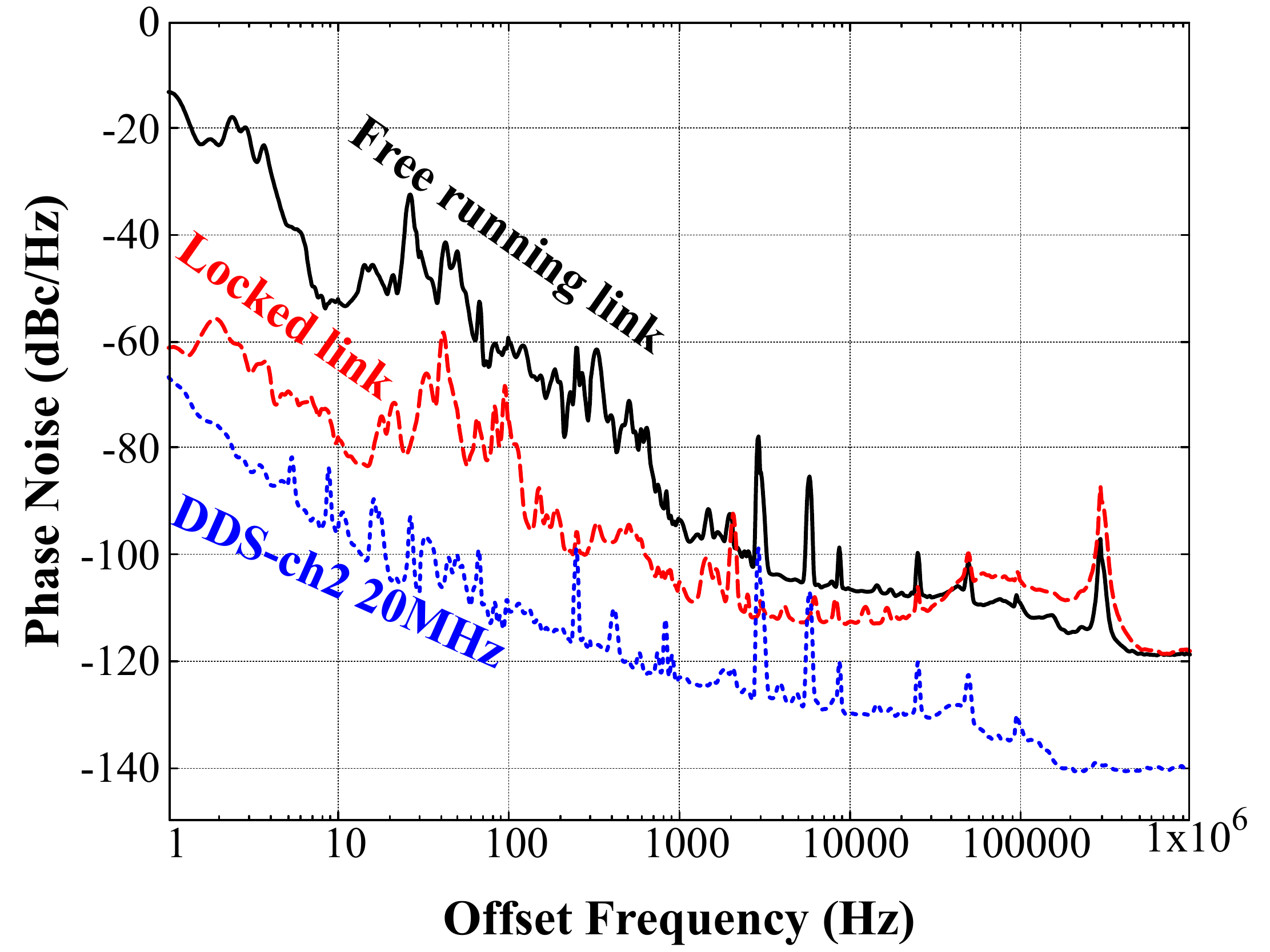} \caption{ }
\label{fig:Phase Noise of Internal Quartz Red Pitaya} 
\end{subfigure}
\caption{Stand-alone digital configuration. (a) Fractional frequency stability
of free running link (solid black circles), locked link (solid red squares), 55
MHz RP output (solid purple triangles), 20~MHz output (solid
inverted blue triangles). (b) Phase Noise of  free running link (black straight
line), locked link (red dashed line), 20~MHz output (blue dotted line). All
RP signals are normalized to the 194~THz carrier frequency.}
\label{fig:RPINT_stab} \end{figure}

In this section, we use the board internal quartz oscillator to clock the digital controls. The results obtained are shown in Fig. \ref{fig:RPINT_stab}.

Fig. \ref{fig:RPINT_stab} (a) shows the fractional frequency stability (FFS, estimated by the modified Allan deviation) of the free-running (black solid circles) and locked (red solid squares) PD2 signal. The locked link stability starts at around $8{\times}10^{-18}$ at 1~s with an initial $\tau^{1/2}$ slope, followed by a $3{\times}10^{-17}$ plateau until 1000~s and a linear frequency drift for longer \textcolor{black}{integration times}. The correction provides more than a factor of 10 instability reduction between 1~s and 1000~s, while the link instability is reduced by a factor 50 at 1~s.

In order to determine the contributions of the digital control board to the link performances, we have also monitored the NCO phase fluctuations with the DDS
implemented on the channel 2 of the DAC set to 55~MHz for free-running measurements and to 20~MHz when the lock is engaged. In open-loop, the 55~MHz frequency noise will be multiplied by 2 at the board output. In closed-loop, the 20~MHz NCO frequency fluctuations are divided by 2 through the control loop at the user end (see \ref{sec:ppe}). To analyze their contributions, we scale the 55~MHz and 20~MHz signals by multiplying their frequency fluctuations by 2 and 0.5, respectively, and normalize them by 194~THz for FFS plots.

The FFS of the $55\ \mathrm{MHz}$
signal shows that the DDS used to drive the AOM is limiting for integration
times longer than $200\ \mathrm{s}$ when the link is free running. This phase
fluctuation is cancelled when the loop is closed and we see that the FFS of a
$20\ \mathrm{MHz}$ signal scaled to the link output frequency overlaps with the
stabilized link FFS indicating that the phase reference of the PLL $\varphi_{0}$
is the limitation. The corresponding DDS FFS at 20~MHz is around
$1.6\times10^{-10}$ at 1~s, consistent with the measurement presented in
\cite{Olaya2019}.

Fig. \ref{fig:RPINT_stab} (b) shows the single-sideband phase noise spectrum of
the free-running (black line) and locked (red dashed line) PD2 signal. The
locked link instability is $-60\ \mathrm{dBc/Hz}$ at $1\ \mathrm{Hz}$, reaching
a $-110\ \mathrm{dBc/Hz}$ floor around $10\ \mathrm{kHz}$. The lock bandwidth is
\textcolor{black}{$\approx40\ \mathrm{kHz}$}.

Please note that the NCO contribution at the link output is of about $-66\ \mathrm{dBc/Hz}$ at 1~Hz. If we scale this back to the digital board clock frequency, this corresponds to $-31 \ \mathrm{dBc/Hz}$ at 1~Hz at 125~MHz. This is in good agreement with the RP internal quartz
phase noise presented in \cite{Olaya2019}. This implies that the NCO performance
is entirely determined by the board internal quartz, justifying the use of an
external clock for improved performance.




\subsection{Using the board clocked by an external clock} \label{sec:RPext}

\begin{figure}[h!] 
\centering 
\begin{subfigure}{.5\textwidth} \centering
\includegraphics[width=8cm]{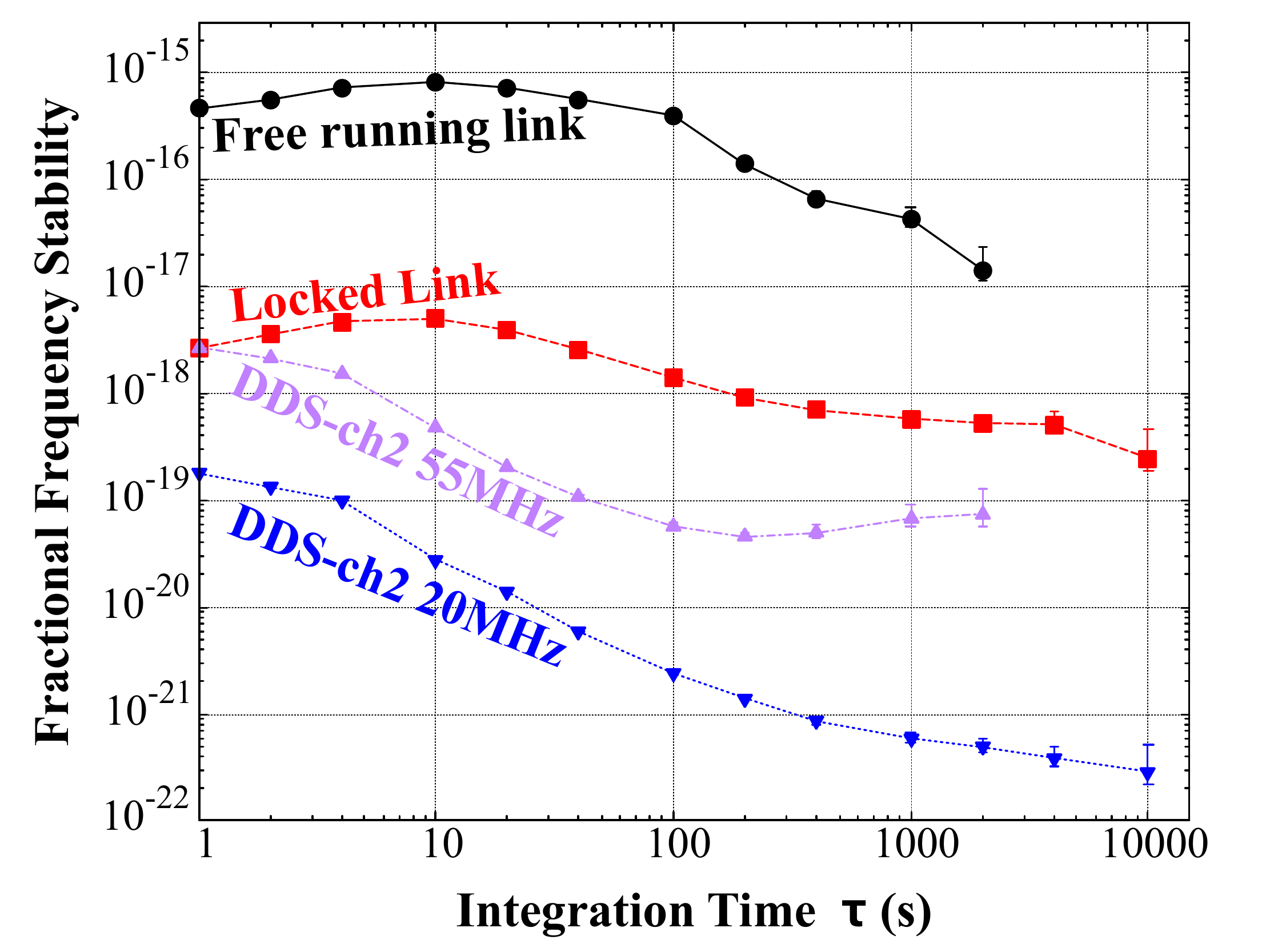} 
\caption{}
\label{fig:suba1} 
\end{subfigure}%
\\ 
\begin{subfigure}{.5\textwidth} \centering
\includegraphics[width=8cm]{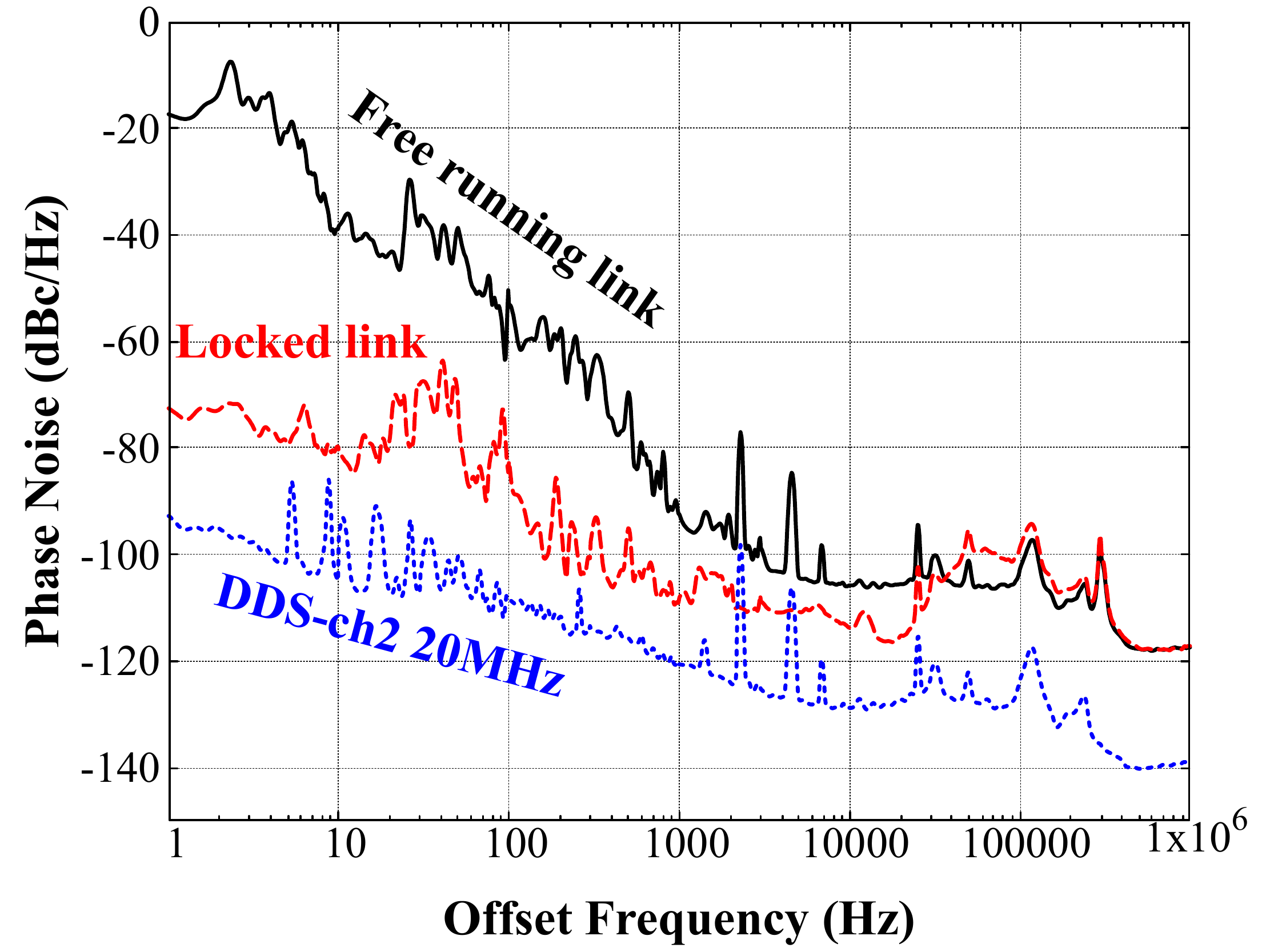} 
\caption{}
\label{fig:suba2} 
\end{subfigure} 
\caption{External clock digital board
configuration. (a) Fractional frequency stability of free running link (solid
black circles), locked link (solid red squares), 55~MHz RP output (solid purple
triangles), 20~MHz output (solid inverted blue triangles). (b)
Phase Noise of  free running link (black straight line), locked link (red dashed
line), 20~MHz output (blue dotted line). All RP signals are normalized to
the 194~THz carrier frequency.} \label{fig:RPEXT_stab} \end{figure}

To further investigate the fundamental limitation of the RP we provide an
external low noise 125~MHz clock signal to the digital control board. This
signal is generated by a DDS clocked by an active \textcolor{black}{HM}, and is coupled to the
board through a balun to respect impedance matching. The contribution of the
clock signal is now estimated at $10^{-19}$ or below. The results are shown in
Fig. \ref{fig:RPEXT_stab}.

Fig. \ref{fig:RPEXT_stab} (a) shows the fractional frequency stability
(estimated by the modified Allan deviation) of the free-running (black solid
circles) and locked (red solid squares) PD2 signal when using an external
clock. The locked link stability starts at around $2{\times}10^{-18}$ at 1~s
with a $8{\times}10^{-18}$ bump at ten seconds, and then reaches a floor below
$10^{-18}$. The link instability is reduced by a factor 100 from 1~s to 400~s.

Fig. \ref{fig:RPEXT_stab} (b) shows the single-sideband phase noise spectrum of
the free-running (black line) and locked (red dashed line) PD2 signal when using
an external clock. The locked link phase noise is below -70~dBc/Hz at 1~Hz,
reaching a $-110 \ \mathrm{dBc/Hz}$ floor around 10~kHz. The lock bandwidth is $\approx40~\rm{kHz}$, as in the previous section. The NCO contribution is again consistent with the external
clock phase noise measured by \cite{Olaya2019}.

\subsection{Implementing undersampling with the Red Pitaya} \label{sec:unders}

In this section, we use undersampling in order to directly extract the phase
noise from the 220~MHz output of the lock photodiode PD1. To do so, we have
patched the commercial board by bypassing the anti-aliasing filters to directly
send the signal to the sample and hold of the ADC \textcolor{black}{(this would be unnecessary for boards such as the STEMlab 122-16 with direct RF sampling up to 550~MHz). The board unbalanced to balanced amplifier that acts as a low pass filter is replaced by a passive balun transformer, so as to benefit from the full 750~MHz bandwidth of the LTC2145 analog to digital converter frontend. This enables undersampling the signal, under the strong assumption of a pure and known beatnote frequency; this is mandatory for recovering the useful signal when it does not comply with the Nyquist–Shannon theorem.}

\begin{figure}[h!] \centering
\includegraphics[width=.5\textwidth]{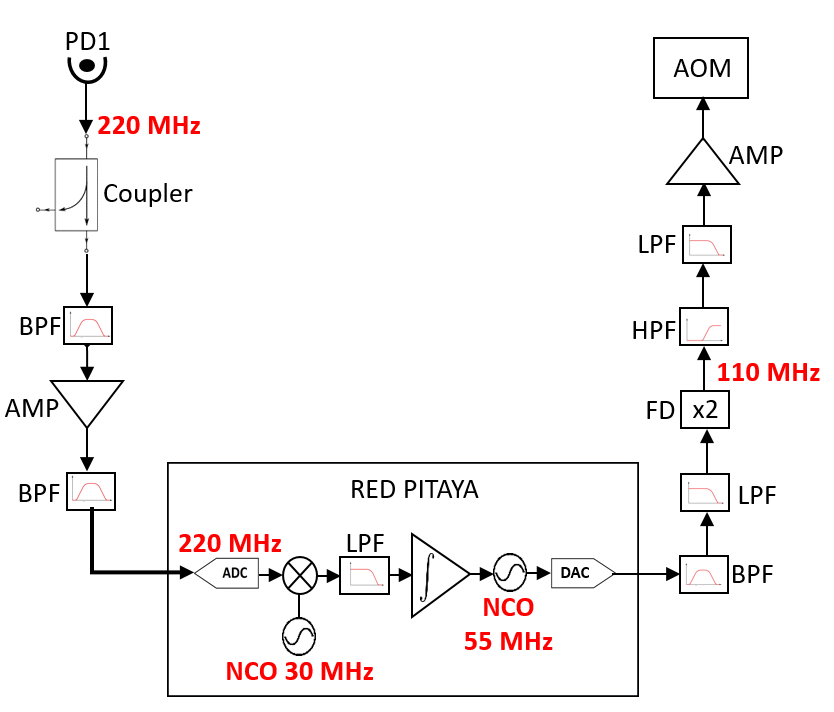}
\caption{Control electronics, undersampling servo loop setup. Fluctuations
carried by the 220~MHz signal are directly sampled by a Red Pitaya to implement
a PLL. Corrections are carried by the 110~MHz signal driving the AOM. See text
for details. \newline PD: Photodiode, LPF: Low Pass Filter, HPF: High Pass
Filter, BPF: Band Pass Filter, AMP: Amplifier, FD: Frequency Doubler. \textcolor{black}{The coupler is for monitoring purposes only.}}
\label{fig:RP_unders_scheme} \end{figure}

We use this method to avoid the use of a mixer and synthesizer at the output of PD1. In addition to removing these potential noise and aging sources, the local oscillator phase is unknown at startup, whereas the ADC is readily frequency and phase synchronized over a network disseminating time and frequency information. Thus, undersampling allows for coherent acquisition on the large scale of a laboratory. As shown Fig. \ref{fig:RP_unders_scheme}, we filter and amplify the photodiode signal and send it directly to the modified board. With an external clock at 125~MHz, we work in the second Nyquist band of the input ADC. As a result, the 220~MHz signal is read as a $2\times 125-220=30$~MHz sine wave. We demodulate this aliased signal using a 30~MHz NCO. The rest of the setup is unchanged compared to previous sections.

The fractional frequency stability and the phase noise measurements are shown in
Fig. \textcolor{black}{\ref{fig:RPunders_comp} by the dashed curves}.

Fig. \textcolor{black}{\ref{fig:RPunders_comp}} (a) shows the fractional frequency stability (estimated
by the modified Allan deviation) of the free-running (black open circles) and
locked (red open squares) PD2 signal for the undersampling scheme. The locked
link stability starts at around \textcolor{black}{$6{\times}10^{-18}$} at 1 second with a
\textcolor{black}{$4{\times}10^{-18}$} bump at ten seconds, and then reaches a floor below
$10^{-18}$. There is a 1/100  reduction of the link instability from 1~s to
200~s.

Fig. \textcolor{black}{\ref{fig:RPunders_comp}} (b) shows the single-sideband phase noise spectrum of the
free-running (black dashed line) and locked (red dashed line) PD2 signal for the
undersampling scheme. The locked link instability is -70~dBc/Hz at 1~Hz,
reaching a -110~dBc/Hz floor around 10 kHz. The lock bandwidth \textcolor{black}{in this configuration is around 70~kHz}.

\section{Discussion} \label{sec:discussion}

The results obtained with a commercial board-based digital control for a
Doppler-compensation scheme are very promising for a wide range of applications that require the local distribution of an optical frequency reference signal.

The measured locking bandwidth for all configurations is between 40 and 70 kHz. With our link length of about 20~m in total, the one-way optical delay $\tau_{link}$ is about 100~ns, which determines the best achievable, link-delay-limited bandwidth of $1/(8\times\tau_{link})\approx1.25~\rm{MHz}$ (assuming a $\pi/4$ phase margin). The digital board main delay stems from the FIR, and is estimated to be at most 136~ns for our 25-coefficients filter.
Lastly, the analog electronics and AOM contribution to the delay is of about 1~$\mu$s. With this total $1.236\ \mu$s delay, the maximum achievable bandwidth in our configuration is around 100~kHz.

Using a stand-alone board, the fractional stability performance is compatible
with compact ultra-stable cavity setups, which have demonstrated fractional
frequency stabilities in the range of $10^{-15}$ as well as ultrastable FP
cavities with stabilities in the range of $10^{-16}$ and slightly below. It is
however not suitable for the best current FP cavities or atomic clocks. For
applications where the optical signal transmitted through the link would serve
as a microwave photonics source, the measured phase noise is remarkably low.
Assuming perfect division to 10~GHz, the measured phase noise contribution of
the link would be $-155\ \mathrm{dBc/Hz}$ at 1~Hz, with a $-205\ \mathrm{dBc/Hz}$ floor.

Using an external clock pushes the setup performance towards state-of-the-art,
and its residual fractional frequency instability is sufficient to transfer
signals at the level of the best FP-based ultrastable lasers in the low
$10^{-17}$. When using an active HM as a reference, the digital board clock performance will actually be limited by its internal PLL residual noise \cite{Olaya2019}. According to \cite{Olaya2019}, using instead a 10 MHz reference signal would degrade the SoC PLL residual noise by roughly a factor of 2.  Given the typical performance of GPS-disciplined Rb standards, this would still guarantee a link fractional frequency stability well below $10^{-17}$ at 1~s and a negligible NCO contribution at longer integration times.

Our measured stabilized link performance is however not suitable for the transfer of the signal from
state-of-the art optical clocks, with instabilities starting at $10^{-16}$ or
below and integrating as $\tau^{-1/2}$. The measured noise of our setup seems to
show a floor near 1000 s which is close to the performance of these clocks. We
however believe that the measured fractional frequency stability is limited by
our HMI setup rather than the link, and that etalon effects due to the absence
of a second AOM near the FM could explain the measured instability. This could be tested in the future to verify  that the presented setup is compatible with today's best optical atomic
clocks. As our setup is mostly targeted towards compact, less-demanding
applications, we have not tried to improve the setup limit. It is clear that the
digital board itself should not limit the setup performance even for
state-of-the art performance, as the NCO contribution is below $10^{-18}$ at all
\textcolor{black}{integration times} and integrates down to the $10^{-22}$ range. Moreover, the ADC flicker phase noise is about $-109\ \mathrm{dBc/Hz}$ at 1~Hz,  with a floor at $-146\ \mathrm{dBc/Hz}$, even for a 220~MHz input frequency, as shown in \cite{Cardenas-Olaya2017}. This makes the ADC phase noise contribution well below the best phase noise reported here.

\begin{figure}[b!] \centering \begin{subfigure}{.5\textwidth} \centering
\includegraphics[width=8cm]{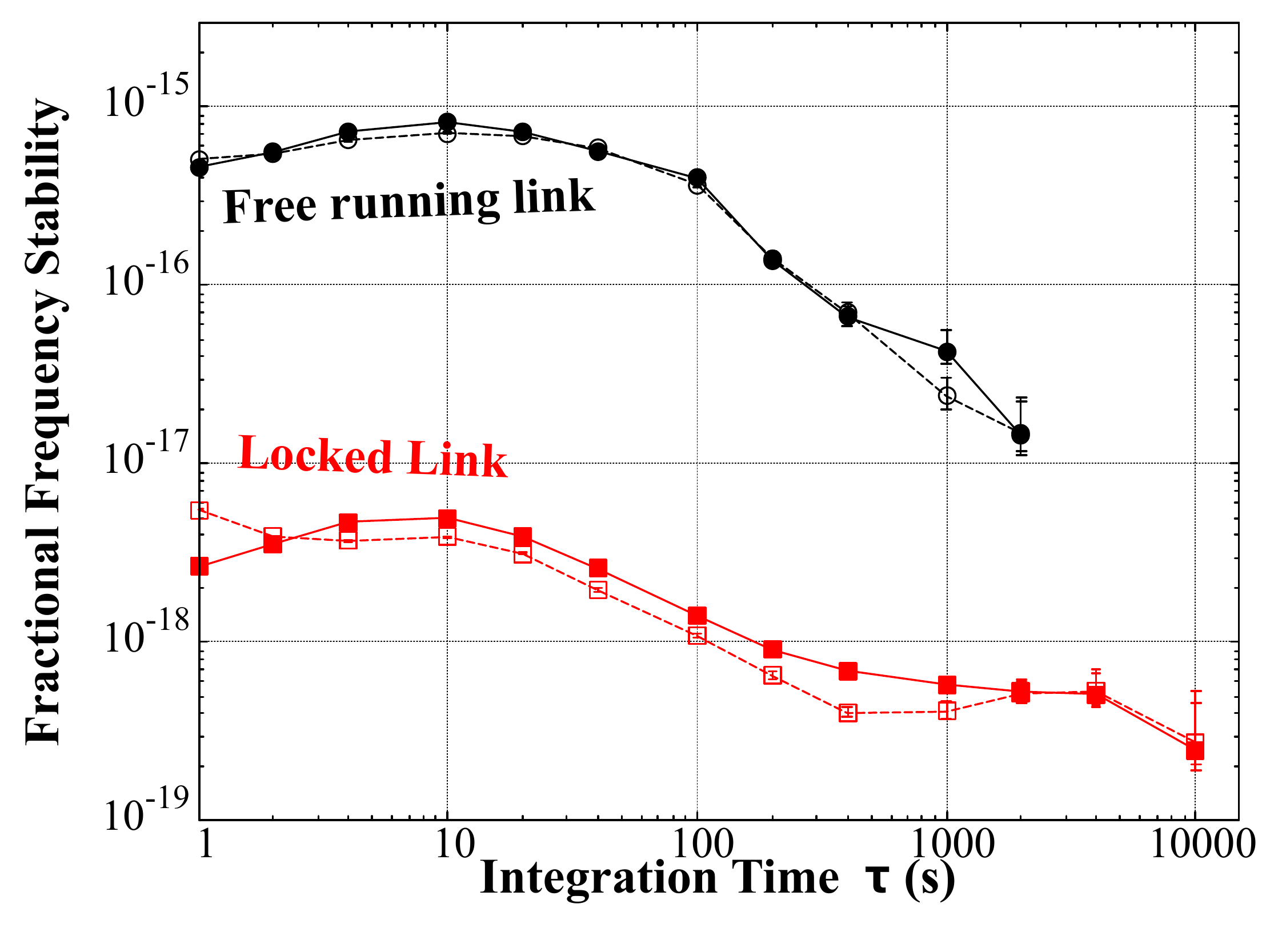} \caption{}
\label{fig:subc2} \end{subfigure}%
\\ \begin{subfigure}{.5\textwidth} \centering
\includegraphics[width=8cm]{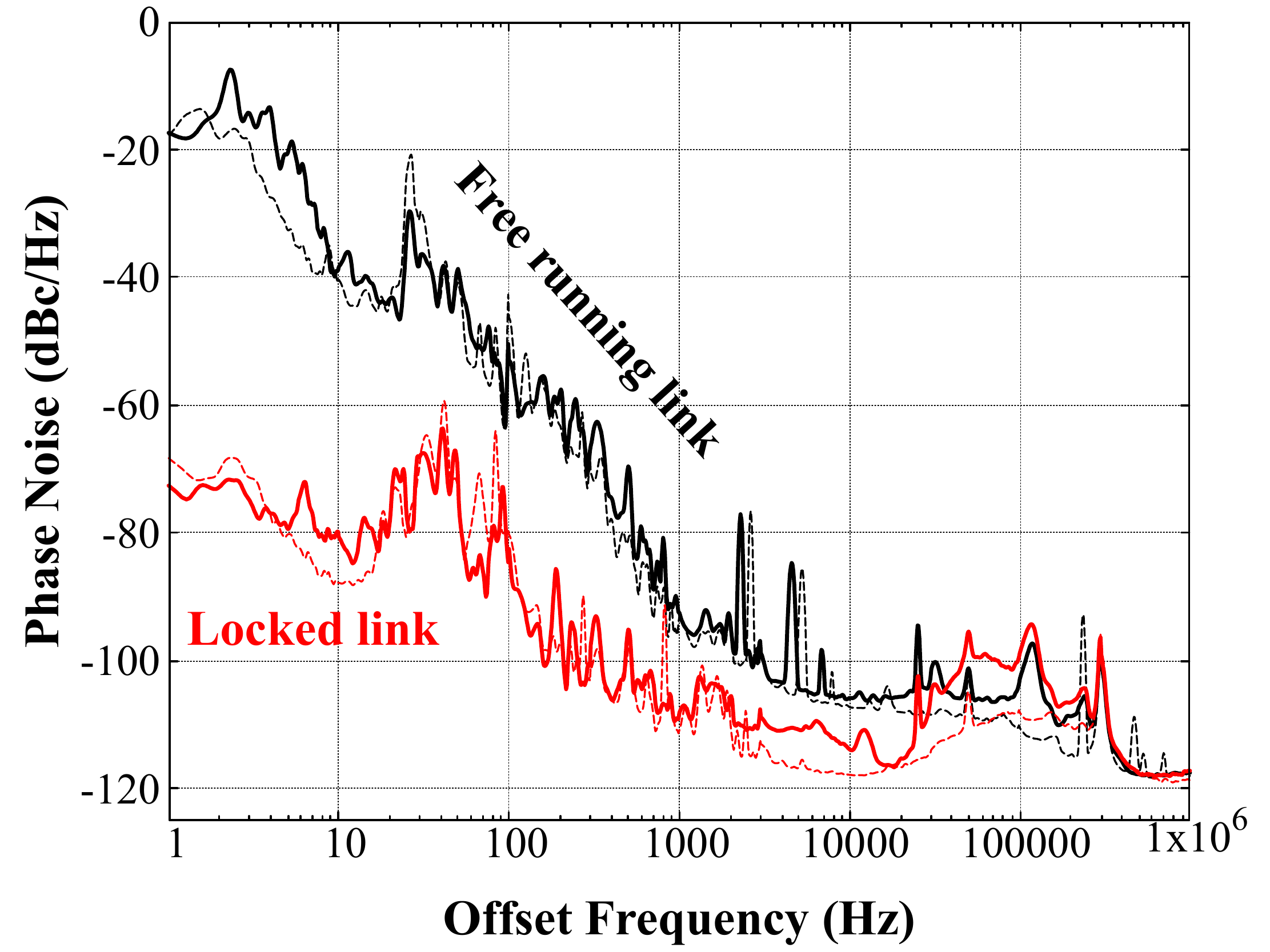} \caption{}
\label{fig:subd2} \end{subfigure}%
\caption{(a) Comparison between the \textcolor{black}{external clock (full symbols)}
and undersampling \textcolor{black}{(with external clock)} configurations, free running link with undersampled Red Pitaya
(open black circles), locked link (open red squares). All RP signals are normalized to the 194~THz carrier
frequency. \newline    (b) Comparison between the Phase Noise of the link with
undersampled Red Pitaya (red dashed, black dashed) and normal Red Pitaya (red
straight, \textcolor{black}{black} straight).} \label{fig:RPunders_comp} 
\end{figure}

As shown \textcolor{black}{in} Fig. \ref{fig:RPunders_comp}, the undersampling setup provides the same
results as the regular externally clocked board, with a simpler setup.

The ch. 2 DDS output is also useful as an indication of the digital board performance. As described in section \ref{sec:RPint}, the in-loop and out-of-loop NCO contributions are scaled differently, by a factor of 0.5 and 2 respectively. Assuming they have identical FFS determined by the board clock FFS, their impact at the optical carrier frequency should differ by a ratio of $4\times55/20=11$. However, we observed that this ratio ranges between 15 and 20 at 1~s for our measurements, while for longer integration times, different behaviors are observed. This seems to imply that the assumption of FFS determined only by the digital board clock fails, and that additional frequency noise is added by the NCO. While this is not a limitation for the purpose presented here, this should be explained and is currently under investigation.

\section{Conclusion}

We have implemented a Doppler-cancellation scheme for optical fiber link phase stabilization that relies on a digital
servo-loop. Using a commercial digital control board in three different
configurations, including using an external clock and undersampling. Using an
external clock, the stabilized link is compatible with current state-of-the-art
optical references. Using undersampling simplifies the setup while maintaining
this level of spectral purity.

While the fractional frequency stability reached with our measurements could be
improved by improving the optical setup, we have determined that the digital
electronic board intrinsic noise would limit the performance to the low
$10^{-22}$ range at 10000 s. This shows that low-cost, small footprint devices
can be integrated to the most demanding metrological setups. All digital designs are available at \newline \url{https://github.com/oscimp/oscimpDigital} and can therefore be easily duplicated.

\section{Acknowledgements} This work has been supported by the EIPHI graduate
school (contract ANR-17-EURE-0002), the labex FIRST-TF (ANR-10-LABX-48-01), the EquipeX OSCILLATOR-IMP (ANR-11-EQPX-0033), the Région Bourgogne Franche-Comté, and the Centre National d'\'Etudes Spatiales (CNES, R\&T R-S19-LN-0001-019).

Séverine Denis is now with CSEM, Rue Jaquet-Droz 1, 2002 Neuchâtel, Switzerland.
Baptiste Marechal is now a Vision Engineer and Developer with Mikron SA, Route du Vignoble 17, 2017 Boudry.

The authors would like to thank Rodolphe Boudot and Marion Delehaye for their careful reading of the manuscript, as well as Léo Tranchart and Ivan Ryger for the FIR delay measurements.
 
\bibliographystyle{IEEEtran}
\typeout{}
\bibliography{bibliographielinks}

\begin{thebibliography}{10}
\providecommand{\url}[1]{#1}
\csname url@samestyle\endcsname
\providecommand{\newblock}{\relax}
\providecommand{\bibinfo}[2]{#2}
\providecommand{\BIBentrySTDinterwordspacing}{\spaceskip=0pt\relax}
\providecommand{\BIBentryALTinterwordstretchfactor}{4}
\providecommand{\BIBentryALTinterwordspacing}{\spaceskip=\fontdimen2\font plus
\BIBentryALTinterwordstretchfactor\fontdimen3\font minus
  \fontdimen4\font\relax}
\providecommand{\BIBforeignlanguage}[2]{{%
\expandafter\ifx\csname l@#1\endcsname\relax
\typeout{** WARNING: IEEEtran.bst: No hyphenation pattern has been}%
\typeout{** loaded for the language `#1'. Using the pattern for}%
\typeout{** the default language instead.}%
\else
\language=\csname l@#1\endcsname
\fi
#2}}
\providecommand{\BIBdecl}{\relax}
\BIBdecl

\bibitem{Haefner2015}
S.~H\"afner, S.~Falke, C.~Grebing, S.~Vogt, T.~Legero, M.~Merimaa, C.~Lisdat,
  and U.~Sterr, ``$8 \times 10^{-17}$ fractional laser frequency instability
  with a long room-temperature cavity,'' \emph{Optics Letters}, vol.~40, no.~9,
  pp. 2112--2115, May 2015.

\bibitem{Matei2017}
D.~Matei, T.~Legero, S.~Häfner, C.~Grebing, R.~Weyrich, W.~Zhang,
  L.~Sonderhouse, J.~Robinson, J.~Ye, F.~Riehle, and U.~Sterr, ``$1.5 \ \mu$m
  {Lasers} with {Sub}-10 {mHz} {Linewidth},'' \emph{Physical Review Letters},
  vol. 118, no.~26, p. 263202, Jun. 2017.

\bibitem{Zhang2017}
W.~Zhang, J.~Robinson, L.~Sonderhouse, E.~Oelker, C.~Benko, J.~Hall, T.~Legero,
  D.~Matei, F.~Riehle, U.~Sterr, and J.~Ye, ``Ultrastable {Silicon} {Cavity} in
  a {Continuously} {Operating} {Closed}-{Cycle} {Cryostat} at 4~{K},''
  \emph{Physical Review Letters}, vol. 119, no.~24, p. 243601, Dec. 2017.

\bibitem{Milner2019}
W.~R. Milner, J.~M. Robinson, C.~J. Kennedy, T.~Bothwell, D.~Kedar, D.~G.
  Matei, T.~Legero, U.~Sterr, F.~Riehle, H.~Leopardi, T.~M. Fortier, J.~A.
  Sherman, J.~Levine, J.~Yao, J.~Ye, and E.~Oelker, ``Demonstration of a
  timescale based on a stable optical carrier,'' \emph{Phys. Rev. Lett.}, vol.
  123, p. 173201, Oct 2019.

\bibitem{Robinson2019}
J.~M. Robinson, E.~Oelker, W.~R. Milner, W.~Zhang, T.~Legero, D.~G. Matei,
  F.~Riehle, U.~Sterr, and J.~Ye, ``Crystalline optical cavity at 4 {K} with
  thermal-noise-limited instability and ultralow drift,'' \emph{Optica},
  vol.~6, no.~2, pp. 240--243, Feb 2019.

\bibitem{Leibrandt2011}
D.~R. Leibrandt, M.~J. Thorpe, J.~C. Bergquist, and T.~Rosenband, ``Field-test
  of a robust, portable, frequency-stable laser,'' \emph{Optics Express},
  vol.~19, no.~11, pp. 10\,278--10\,286, May 2011.

\bibitem{Vogt2011}
S.~Vogt, C.~Lisdat, T.~Legero, U.~Sterr, I.~Ernsting, A.~Nevsky, and
  S.~Schiller, ``Demonstration of a transportable 1 {Hz}-linewidth laser,''
  \emph{Applied Physics B}, vol. 104, no.~4, p. 741, Sep. 2011.

\bibitem{Argence2012}
B.~Argence, E.~Prevost, T.~Lévèque, R.~Le~Goff, S.~Bize, P.~Lemonde, and
  G.~Santarelli, ``Prototype of an ultra-stable optical cavity for space
  applications,'' \emph{Optics Express}, vol.~20, no.~23, pp. 25\,409--25\,420,
  Nov. 2012.

\bibitem{Haefner2020}
S.~Häfner, S.~Häfner, S.~Herbers, S.~Vogt, S.~Vogt, C.~Lisdat, and U.~Sterr,
  ``Transportable interrogation laser system with an instability of mod
  $\sigma_y=3\times10^{-16}$,'' \emph{Optics Express}, vol.~28, no.~11, pp.
  16\,407--16\,416, May 2020, publisher: Optical Society of America.

\bibitem{Lezius2016}
M.~Lezius, T.~Wilken, C.~Deutsch, M.~Giunta, O.~Mandel, A.~Thaller,
  V.~Schkolnik, M.~Schiemangk, A.~Dinkelaker, A.~Kohfeldt, A.~Wicht,
  M.~Krutzik, A.~Peters, O.~Hellmig, H.~Duncker, K.~Sengstock,
  P.~Windpassinger, K.~Lampmann, T.~Hülsing, T.~W. Hänsch, and R.~Holzwarth,
  ``Space-borne frequency comb metrology,'' \emph{Optica}, vol.~3, no.~12, pp.
  1381--1387, Dec. 2016.

\bibitem{Hinkley2013}
N.~Hinkley, J.~A. Sherman, N.~B. Phillips, M.~Schioppo, N.~D. Lemke, K.~Beloy,
  M.~Pizzocaro, C.~W. Oates, and A.~D. Ludlow, ``An {Atomic} {Clock} with
  $10^{–18}$ {Instability},'' \emph{Science}, vol. 341, no. 6151, pp.
  1215--1218, Sep. 2013.

\bibitem{Bloom2014}
B.~J. Bloom, T.~L. Nicholson, J.~R. Williams, S.~L. Campbell, M.~Bishof,
  X.~Zhang, W.~Zhang, S.~L. Bromley, and J.~Ye, ``An optical lattice clock with
  accuracy and stability at the $10^{-18}$ level,'' \emph{Nature}, vol. 506,
  no. 7486, pp. 71--75, Feb. 2014.

\bibitem{Ushijima2015}
I.~Ushijima, M.~Takamoto, M.~Das, T.~Ohkubo, and H.~Katori, ``Cryogenic optical
  lattice clocks,'' \emph{Nature Photonics}, vol.~9, no.~3, pp. 185--189, Mar.
  2015.

\bibitem{Huntemann2016}
N.~Huntemann, C.~Sanner, B.~Lipphardt, C.~Tamm, and E.~Peik, ``Single-{Ion}
  {Atomic} {Clock} with $3 \times 10^{-18}$ {Systematic} {Uncertainty},''
  \emph{Physical Review Letters}, vol. 116, no.~6, p. 063001, Feb. 2016.

\bibitem{Brewer2019}
S.~Brewer, J.-S. Chen, A.~Hankin, E.~Clements, C.~Chou, D.~Wineland, D.~Hume,
  and D.~Leibrandt, ``$^{27}${Al}$^+$ {Quantum}-{Logic} {Clock} with a
  {Systematic} {Uncertainty} below $10^{-18}$,'' \emph{Physical Review
  Letters}, vol. 123, no.~3, p. 033201, Jul. 2019.

\bibitem{Giunta2020}
M.~Giunta, J.~Yu, M.~Lessing, M.~Fischer, M.~Lezius, X.~Xie, G.~Santarelli,
  Y.~L. Coq, and R.~Holzwarth, ``Compact and ultrastable photonic microwave
  oscillator,'' \emph{Optics Letters}, vol.~45, no.~5, pp. 1140--1143, Mar.
  2020, publisher: Optical Society of America.

\bibitem{Derevianko2014}
A.~Derevianko and M.~Pospelov, ``Hunting for topological dark matter with
  atomic clocks,'' \emph{Nature Physics}, vol.~10, no.~12, pp. 933--936, Dec.
  2014.

\bibitem{Safronova2018}
M.~Safronova, D.~Budker, D.~DeMille, D.~F.~J. Kimball, A.~Derevianko, and C.~W.
  Clark, ``Search for new physics with atoms and molecules,'' \emph{Reviews of
  Modern Physics}, vol.~90, no.~2, p. 025008, Jun. 2018, publisher: American
  Physical Society.

\bibitem{Riehle2017}
F.~Riehle, ``Optical clock networks,'' \emph{Nature Photonics}, vol.~11, no.~1,
  pp. 25--31, Jan. 2017.

\bibitem{Lisdat2016}
C.~Lisdat, G.~Grosche, N.~Quintin, C.~Shi, S.~M.~F. Raupach, C.~Grebing,
  D.~Nicolodi, F.~Stefani, A.~Al-Masoudi, S.~Dörscher, S.~Häfner, J.-L.
  Robyr, N.~Chiodo, S.~Bilicki, E.~Bookjans, A.~Koczwara, S.~Koke, A.~Kuhl,
  F.~Wiotte, F.~Meynadier, E.~Camisard, M.~Abgrall, M.~Lours, T.~Legero,
  H.~Schnatz, U.~Sterr, H.~Denker, C.~Chardonnet, Y.~L. Coq, G.~Santarelli,
  A.~Amy-Klein, R.~L. Targat, J.~Lodewyck, O.~Lopez, and P.-E. Pottie, ``A
  clock network for geodesy and fundamental science,'' \emph{Nature
  Communications}, vol.~7, p. 12443, Aug. 2016.

\bibitem{Kudeyarov2018}
K.~S. Kudeyarov, G.~A. Vishnyakova, K.~Y. Khabarova, and N.~N. Kolachevsky,
  ``2.8 km fiber link with phase noise compensation for transportable {Yb}$^+$
  optical clock characterization,'' \emph{Laser Physics}, vol.~28, no.~10, p.
  105103, Jul. 2018, publisher: IOP Publishing.

\bibitem{Beloy2021}
K.~Beloy, M.~I. Bodine, T.~Bothwell, S.~M. Brewer, S.~L. Bromley, J.-S. Chen,
  J.-D. Deschênes, S.~A. Diddams, R.~J. Fasano, T.~M. Fortier, Y.~S. Hassan,
  D.~B. Hume, D.~Kedar, C.~J. Kennedy, I.~Khader, A.~Koepke, D.~R. Leibrandt,
  H.~Leopardi, A.~D. Ludlow, W.~F. McGrew, W.~R. Milner, N.~R. Newbury,
  D.~Nicolodi, E.~Oelker, T.~E. Parker, J.~M. Robinson, S.~Romisch, S.~A.
  Schäffer, J.~A. Sherman, L.~C. Sinclair, L.~Sonderhouse, W.~C. Swann,
  J.~Yao, J.~Ye, X.~Zhang, and {Boulder Atomic Clock Optical Network (BACON)
  Collaboration*}, ``Frequency ratio measurements at 18-digit accuracy using an
  optical clock network,'' \emph{Nature}, vol. 591, no. 7851, pp. 564--569,
  Mar. 2021.

\bibitem{Newbury2007}
N.~R. Newbury, P.~A. Williams, and W.~C. Swann, ``Coherent transfer of an
  optical carrier over 251 km,'' \emph{Optics Letters}, vol.~32, no.~21, pp.
  3056--3058, Nov. 2007.

\bibitem{Bercy2014b}
A.~Bercy, F.~Stefani, O.~Lopez, C.~Chardonnet, P.-E. Pottie, and A.~Amy-Klein,
  ``Two-way optical frequency comparisons at $5\times 10^{-21}$ relative
  stability over 100-km telecommunication network fibers,'' \emph{Physical
  Review A}, vol.~90, no.~6, p. 061802, Dec. 2014, publisher: American Physical
  Society.

\bibitem{Cantin2021}
E.~Cantin, M.~T{\o}nnes, R.~L. Targat, A.~Amy-Klein, O.~Lopez, and P.-E.
  Pottie, ``An accurate and robust metrological network for coherent optical
  frequency dissemination,'' \emph{New Journal of Physics}, vol.~23, no.~5, p.
  053027, may 2021.

\bibitem{Lopez2012}
O.~Lopez, A.~Haboucha, B.~Chanteau, C.~Chardonnet, A.~Amy-Klein, and
  G.~Santarelli, ``Ultra-stable long distance optical frequency distribution
  using the {Internet} fiber network,'' \emph{Optics Express}, vol.~20, no.~21,
  pp. 23\,518--23\,526, Oct. 2012.

\bibitem{Calonico2014a}
D.~Calonico, E.~K. Bertacco, C.~E. Calosso, C.~Clivati, G.~A. Costanzo,
  M.~Frittelli, A.~Godone, A.~Mura, N.~Poli, D.~V. Sutyrin, G.~Tino, M.~E.
  Zucco, and F.~Levi, ``High-accuracy coherent optical frequency transfer over
  a doubled 642-km fiber link,'' \emph{Applied Physics B}, vol. 117, no.~3, pp.
  979--986, Dec. 2014.

\bibitem{Lopez2015}
O.~Lopez, F.~Kéfélian, H.~Jiang, A.~Haboucha, A.~Bercy, F.~Stefani,
  B.~Chanteau, A.~Kanj, D.~Rovera, J.~Achkar, C.~Chardonnet, P.-E. Pottie,
  A.~Amy-Klein, and G.~Santarelli, ``Frequency and time transfer for metrology
  and beyond using telecommunication network fibres,'' \emph{Comptes Rendus
  Physique}, vol.~16, no.~5, pp. 531--539, Jun. 2015.

\bibitem{Clivati2020}
C.~Clivati, R.~Aiello, G.~Bianco, C.~Bortolotti, P.~D. Natale, V.~D. Sarno,
  P.~Maddaloni, G.~Maccaferri, A.~Mura, M.~Negusini, F.~Levi, F.~Perini,
  R.~Ricci, M.~Roma, L.~S. Amato, M.~S. de~Cumis, M.~Stagni, A.~Tuozzi, and
  D.~Calonico, ``Common-clock very long baseline interferometry using a
  coherent optical fiber link,'' \emph{Optica}, vol.~7, no.~8, pp. 1031--1037,
  Aug 2020.

\bibitem{Ma1994}
L.-S. Ma, P.~Jungner, J.~Ye, and J.~L. Hall, ``Delivering the same optical
  frequency at two places: accurate cancellation of phase noise introduced by
  an optical fiber or other time-varying path,'' \emph{Optics Letters},
  vol.~19, no.~21, pp. 1777--1779, Nov. 1994.

\bibitem{Ye2003}
J.~Ye, J.-L. Peng, R.~J. Jones, K.~W. Holman, J.~L. Hall, D.~J. Jones, S.~A.
  Diddams, J.~Kitching, S.~Bize, J.~C. Bergquist, L.~W. Hollberg,
  L.~Robertsson, and L.-S. Ma, ``Delivery of high-stability optical and
  microwave frequency standards over an optical fiber network,'' \emph{JOSA B},
  vol.~20, no.~7, pp. 1459--1467, Jul. 2003.

\bibitem{Calosso2015}
C.~E. Calosso, E.~K. Bertacco, D.~Calonico, C.~Clivati, G.~A. Costanzo,
  M.~Frittelli, F.~Levi, S.~Micalizio, A.~Mura, and A.~Godone,
  ``Doppler-stabilized fiber link with 6{dB} noise improvement below the
  classical limit,'' \emph{Optics Letters}, vol.~40, no.~2, pp. 131--134, Jan.
  2015.

\bibitem{Stefani2015}
F.~Stefani, O.~Lopez, A.~Bercy, W.-K. Lee, C.~Chardonnet, G.~Santarelli, P.-E.
  Pottie, and A.~Amy-Klein, ``Tackling the limits of optical fiber links,''
  \emph{JOSA B}, vol.~32, no.~5, pp. 787--797, May 2015.

\bibitem{Raupach2015}
S.~M.~F. Raupach, A.~Koczwara, and G.~Grosche, ``Brillouin amplification
  supports $1\times10^{-20}$ uncertainty in optical frequency transfer over
  1400 km of underground fiber,'' \emph{Physical Review A}, vol.~92, no.~2, p.
  021801, Aug. 2015.

\bibitem{Chiodo2015a}
N.~Chiodo, N.~Quintin, F.~Stefani, F.~Wiotte, E.~Camisard, C.~Chardonnet,
  G.~Santarelli, A.~Amy-Klein, P.-E. Pottie, and O.~Lopez, ``Cascaded optical
  fiber link using the internet network for remote clocks comparison,''
  \emph{Optics Express}, vol.~23, no.~26, pp. 33\,927--33\,937, Dec. 2015.

\bibitem{Bowler2013}
R.~Bowler, U.~Warring, J.~W. Britton, B.~C. Sawyer, and J.~Amini, ``Arbitrary
  waveform generator for quantum information processing with trapped ions,''
  \emph{Review of Scientific Instruments}, vol.~84, no.~3, p. 033108, Mar.
  2013, publisher: American Institute of Physics.

\bibitem{Huang2014}
K.~Huang, H.~Le~Jeannic, J.~Ruaudel, O.~Morin, and J.~Laurat,
  ``Microcontroller-based locking in optics experiments,'' \emph{Review of
  Scientific Instruments}, vol.~85, no.~12, p. 123112, Dec. 2014.

\bibitem{Pruttivarasin2015}
T.~Pruttivarasin and H.~Katori, ``Compact field programmable gate array-based
  pulse-sequencer and radio-frequency generator for experiments with trapped
  atoms,'' \emph{Review of Scientific Instruments}, vol.~86, no.~11, p. 115106,
  Nov. 2015, publisher: American Institute of Physics.

\bibitem{Stuart2019}
J.~Stuart, R.~Panock, C.~Bruzewicz, J.~Sedlacek, R.~McConnell, I.~Chuang,
  J.~Sage, and J.~Chiaverini, ``Chip-{Integrated} {Voltage} {Sources} for
  {Control} of {Trapped} {Ions},'' \emph{Physical Review Applied}, vol.~11,
  no.~2, p. 024010, Feb. 2019, publisher: American Physical Society.

\bibitem{sdra}
E.~Richter, ``Usage of higher order nyquist zones with direct sampling
  devices,'' in \emph{Software Defined Radio Academy (SDRA)}, 2020,
  \url{https://www.youtube.com/watch?v=PI_ROLXqO_Q}.

\bibitem{hasselwander2016gr}
C.~J. Hasselwander, Z.~Cao, and W.~A. Grissom, ``gr-mri: A software package for
  magnetic resonance imaging using software defined radios,'' \emph{Journal of
  Magnetic Resonance}, vol. 270, pp. 47--55, 2016.

\bibitem{sherman2016oscillator}
J.~A. Sherman and R.~J{\"o}rdens, ``Oscillator metrology with software defined
  radio,'' \emph{Review of Scientific Instruments}, vol.~87, no.~5, p. 054711,
  2016.

\bibitem{mahnke2018characterization}
P.~Mahnke, ``Characterization of a commercial software defined radio as high
  frequency lock-in amplifier for fm spectroscopy,'' \emph{Review of Scientific
  Instruments}, vol.~89, no.~1, p. 013113, 2018.

\bibitem{balakirev2019resonant}
F.~F. Balakirev, S.~M. Ennaceur, R.~J. Migliori, B.~Maiorov, and A.~Migliori,
  ``Resonant ultrasound spectroscopy: The essential toolbox,'' \emph{Review of
  Scientific Instruments}, vol.~90, no.~12, p. 121401, 2019.

\bibitem{tourigny2018open}
A.~Tourigny-Plante, V.~Michaud-Belleau, N.~Bourbeau~H{\'e}bert, H.~Bergeron,
  J.~Genest, and J.-D. Desch{\^e}nes, ``An open and flexible digital
  phase-locked loop for optical metrology,'' \emph{Review of Scientific
  Instruments}, vol.~89, no.~9, p. 093103, 2018.

\bibitem{stimpson2019open}
G.~Stimpson, M.~Skilbeck, R.~Patel, B.~Green, and G.~Morley, ``An open-source
  high-frequency lock-in amplifier,'' \emph{Review of Scientific Instruments},
  vol.~90, no.~9, p. 094701, 2019.

\bibitem{preuschoff2020digital}
T.~Preuschoff, M.~Schlosser, and G.~Birkl, ``Digital laser frequency and
  intensity stabilization based on the stemlab platform (originally red
  pitaya),'' \emph{Review of Scientific Instruments}, vol.~91, no.~8, p.
  083001, 2020.

\bibitem{Deschenes}
J.-D. Deschenes, digital PLL code base, see
  https://github.com/jddes/Frequency-comb-DPLL.

\bibitem{Olaya2021}
\BIBentryALTinterwordspacing
C.~Cardenas-Olaya and C.~Calosso, ``\BIBforeignlanguage{en}{Fully digital
  electronics for fiber-link frequency transfer implemented on red pitaya},''
  \emph{\BIBforeignlanguage{en}{Proceedings of the GNU Radio Conference}},
  vol.~2, no.~1, Jan. 2021. [Online]. Available:
  \url{https://pubs.gnuradio.org/index.php/grcon/article/view/95}
\BIBentrySTDinterwordspacing

\bibitem{Olaya2016}
C.~Olaya, A.~C., S.~Micalizio, M.~Ortolano, C.~E. Calosso, E.~Rubiola, and
  J.-M. Friedt, ``Digital electronics based on red pitaya platform for coherent
  fiber links,'' Apr. 2016, pp. 1--4.

\bibitem{Olaya2019a}
A.~C. Cárdenas-Olaya, A.~Tampellini, E.~Bertacco, C.~Clivati, A.~Mura,
  D.~Calonico, S.~Micalizio, and C.~E. Calosso, ``Digital instrumentation for
  phase-coherent frequency transfer over 300 km fiber link,'' Apr. 2019, pp.
  1--2.

\bibitem{Didier2018}
A.~Didier, J.~Millo, B.~Marechal, C.~Rocher, E.~Rubiola, R.~Lecomte, M.~Ouisse,
  J.~Delporte, C.~Lacroûte, and Y.~Kersalé, ``Ultracompact reference ultralow
  expansion glass cavity,'' \emph{Applied Optics}, vol.~57, no.~22, pp.
  6470--6473, Aug. 2018.

\bibitem{cecil}
See http://cecill.info/index.en.html and
  {https://spdx.org/licenses/CECILL-B.html\#licenseText}.

\bibitem{williams2008}
P.~A. Williams, W.~C. Swann, and N.~R. Newbury,
  ``\BIBforeignlanguage{EN}{High-stability transfer of an optical frequency
  over long fiber-optic links},'' \emph{\BIBforeignlanguage{EN}{JOSA B}},
  vol.~25, no.~8, pp. 1284--1293, Aug. 2008.

\bibitem{Olaya2019}
A.~C.~C. Olaya, C.~E. Calosso, J.-M. Friedt, S.~Micalizio, and E.~Rubiola,
  ``Phase {Noise} and {Frequency} {Stability} of the {Red}-{Pitaya} {Internal}
  {PLL},'' \emph{IEEE Transactions on Ultrasonics, Ferroelectrics, and
  Frequency Control}, vol.~66, no.~2, pp. 412--416, Feb. 2019.

\bibitem{Cardenas-Olaya2017}
A.~C. Cárdenas-Olaya, E.~Rubiola, J.-M. Friedt, P.-Y. Bourgeois, M.~Ortolano,
  S.~Micalizio, and C.~E. Calosso, ``Noise characterization of analog to
  digital converters for amplitude and phase noise measurements,'' \emph{Review
  of Scientific Instruments}, vol.~88, no.~6, p. 065108, Jun. 2017.

\end{thebibliography}

\end{document}